# Proceedings for the 28th Annual Conference of the Society for Astronomical Sciences

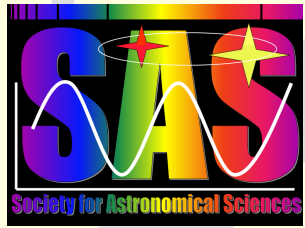
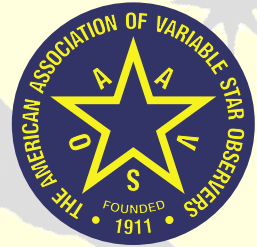

## Joint Meeting of
## The Society for Astronomical Sciences
## The American Association of Variable Star Observers

# Symposium on Telescope Science

Editors:
Brian D. Warner
Jerry Foote
David A. Kenyon
Dale Mais

May 19-21, 2009
Big Bear Lake, CA

# Sloan-r' photometry of Comet 17P/Holmes beyond 3.8 AU: An Observing Methodology for Short-period Comets Far from Perihelion

*Richard Miles*
*Golden Hill Observatory*
*Stourton Caundle, Dorset DT10 2JP, United Kingdom*
*rmiles@baa.u-net.com*

**Abstract**

A photometric method is described for accurately quantifying the brightness of short-period comets far from perihelion. The method utilizes the Sloan Digital Sky Survey Catalog (Data Release 7) as a homogeneous source of reference star magnitudes. Results are based on SDSS-r' filtered images taken using 2.0-m aperture telescopes for which the exposure time was adjusted to achieve a constant motion-blur of 2.0 pixels (0.56 arcsec) on the CCD chip. Aperture photometry using circular and tilted elliptical apertures was performed on images, which were stacked to increase signal to noise. Magnitude dependence on 'seeing' was determined, and this calibration was used to normalize photometry to constant seeing thereby maximizing photometric accuracy. From observations of Comet 17P/Holmes between 2008 October and 2009 March, a very significant outburst of 17P was found to have occurred on 2009 Jan 4.7 (±0.5 day). Night-to-night measurements of the brightness of the inner coma (3000-km radius) exhibited a scatter of only 0.015-0.019 mag. No short time-scale (<36 hr) periodicity was found in the fading lightcurve. From literature data, it was estimated that reflected light from the nucleus contributed 7-11% of the signal within the inner coma and it is concluded that either the nucleus of 17P must be relatively spherical (projected axial ratio of <1.25), or, if its shape is more typical of other comet nuclei, it has a rotational period in excess of 10 days (assuming the observations were not made with the nucleus 'pole-on' to the Earth). Evidence from intermittent activity displayed by the nucleus is indicative of a possible 44 day rotation period.

## 1. Introduction

Comets comprise a wide variety of objects, the orbits of which have been perturbed such that they pass close by the sun, are heated and subsequently emit dust and gas to form a visible coma or tail. Short-period comets make repeated passes through perihelion losing gas and dust each time but as they recede from perihelion they quickly cool and decline in activity. Photometry of active comets can be used to estimate the rate of dust production emitted from the nucleus. The most widely used approach is based on the model of A'Hearn and colleagues (1984) whereby an assumption is made that dust is ejected isotropically and travels away from the nucleus at constant velocity. This model predicts that the integrated magnitude within a circular aperture centered on the pseudo-nucleus is proportional to the radius, $\rho$, of the photometric aperture. One key feature of the model is the 'Afrho' parameter defined as:

$$A f \rho = (2Dr/\rho)^2 \cdot (F_{com}/F_{sol}) \cdot \rho \qquad (1)$$

where, A is the albedo, f is the filling factor of the grains within the photometric aperture, D is the comet-Earth distance in AU, r is the heliocentric distance of the comet in AU, $F_{com}$ is the observed flux reflected within the photometric aperture and $F_{sol}$ is the flux of the Sun at 1 AU. This parameter equates to the diameter of a hypothetical disk of dust, which reflects the same amount of light as the dust in the coma. By use of the isotropic flow model and making assumptions about the density, size and velocity distribution of dust grains, the Afrho quantity can be related directly to the dust production rate at the nucleus (Kidger, 2004). This approach has been very successful in interpreting processes in active comets where dust production takes place from active sites representing a small fraction of the surface of the nucleus. However, for post-perihelion comets, temperatures decrease significantly with increasing heliocentric distance, and the concept of continual dust production may not accurately represent conditions at the nucleus if active regions shut down.

Since dust is lifted away from the surface by viscous drag from escaping gases, the grains will possess a wide velocity distribution and so even if gas production were to suddenly shut down, some dust grains would remain near the nucleus, slowly expanding into space and dissipating with time. This would create the illusion of a declining gas production rate whereas what we may be observing is a solid comet nucleus, akin in properties to those of an asteroid surrounded by a slowly expanding dust cloud dispersal of which is aided by the solar wind. When a Jupiter-family comet is say more than 18 months past perihelion, an alternative model may be valid; that of an inactive rotating nucleus embedded in a cloud of dust grains. Variable aperture photometry can be performed on images of the nucleus and coma, for example to look for any periodic modulation in the brightness of the very inner coma characteristic of the reflected light contribution from a rotating nucleus. Since comet nuclei tend to be significantly more elongated than asteroids, it is feasible to look for rotational modulation even when a substantial inner coma is still present. The challenge is largely one of carrying out aperture photometry to a high degree of precision and accuracy.

## 2. Existing Methodologies for Comet Photometry

### 2.1 ICQ – Visual and CCD photometry

The traditional approach to the photometry of comets is set out in the note on this subject at the International Comet Quarterly (ICQ) website: *http://www.cfa.harvard.edu/icq/cometphot.html*. The format for reporting observations has developed over the years based on visual and photographic observations of comets. In recent years the ICQ format has been adapted to include reports based on CCD observations although magnitudes are generally only reported to a precision of the nearest one decimal place (at most two places of decimal can be reported using this format). Magnitude reports are generally quoted for a single photometric aperture. No reference is made in the ICQ webpages to the Sloan Digital Sky Survey (SDSS) catalogs as a source of magnitude data, nor to the SDSS photometric system.

An article entitled, *CCD Photometry of Comets* can also be found on the ICQ website at *http://www.cfa.harvard.edu/icq/CCDmags.html*, but this has limited relevance as it was written in 1997 since when technology and observing methods have changed considerably.

### 2.2 Cometary Archive for Amateur Astronomers CARA

This project was developed by the Comet Section of the *Unione Astrofili Italiani* (Italian Amateur Astronomer Union) arising from collaboration between active amateur and professional astronomers. Details can be found at *http://www.cara-project.org/*. The main goal of CARA is to collect good photometric data based on the Afrho photometric quantity. A Windows-based computer program, WinAfrho (currently version 1.85), has been developed to assist in the photometric reduction of CCD images. The methodology includes the use of R and I filters along with Hipparcos stars to improve calibration accuracy.

### 2.3 *FOCAS II* and the Spanish Approach

The most significant recent advance in the photometry of comets is the result of the work of the Spanish amateur, Julio Castellano, and his colleagues in developing advanced software named *FOCAS* for automating astrometric and photometric reduction of CCD images. The title is an acronym of '**FO**tometría **C**on **AS**trometrica'. Details can be found at *http://astrosurf.com/cometas-obs/_Articulos/Focas/Focas.htm*. The approach used is to carry out multi-aperture photometry using square aperture boxes (multi-box photometry) centered on the pseudo-nucleus. The first version of the software was published in December 2003 and has undergone numerous revisions since. It uses the software, *Astrometrica*, written by Herbert Raab, to derive a plate solution for each image analysed. A star subtraction routine is employed to isolate the contribution from the coma. One very recent innovation has been to include the *Carlsberg Meridian Catalogue* (CMC-14) as a source of reference star magnitudes. The latest version is dated 2009 Jan 24. *FOCAS II* software supports both Spanish and English language and also permits the photometric analysis of asteroids and variable stars.

Several dozen Spanish amateur astronomers and others employ *FOCAS II* in their work. Many of the active amateurs subscribe to the Yahoo e-group, entitled *Observadores_cometas*, at *http://tech.groups.yahoo.com/group/Observadores_cometas/*. *FOCAS II* generates astrometric reports in the standard Minor Planet Center (MPC) format, along with a special format for reporting photometric results with magnitudes quoted to the nearest two decimal places along with an estimate of the error involved in their determination and other related data. The results on comets are compiled at *http://astrosurf.com/cometas-obs/*.

## 3. 2007 Outburst of Comet 17P/Holmes

### 3.1 Orbital characteristics

As with other Jupiter-family comets, the orbit of 17P is subject to significant change following rela-tively close approaches to Jupiter. For the 2007 return, the comet traveled from an aphelion distance of 5.19 AU reaching a perihelion of 2.05 AU on 2007 May 4. The orbital inclination is 19.1 degrees and the orbital period is 6.9 years. Its perihelion distance at its previous return was 2.18 AU.

### 3.2 Super-outburst of October 23/24

The outburst of Comet Holmes on 2007 Oct 23.7 ± 0.2 d is unrivalled as the most powerful event of its kind on record (Sekanina, 2008). It took place almost 6 months after perihelion passage when the comet had reached 2.44 AU from the Sun. The only other time this object has been known to undergo outbursts was on 1892 Nov 4 (a few days before its discovery by Edwin Holmes) and on 1893 Jan 16, when it was some 5 months, and 7 months, respectively post-perihelion. The amplitude of the 2007 super-outburst amounted to 14 magnitudes, peak brightness being attained almost 2 days after the initial explosion. Sekanina (2008) estimated that about 2% mass was lost from the nucleus.

### 3.3 The Subsequent Fade and a Possible 45-day Periodicity in Activity of the Comet Nucleus

The total magnitude of the near-spherical dust coma remained more or less constant at magnitude 2.5-3.0 for several weeks although it continued to expand during this time at a speed of about 0.5 km/s. However, the inner coma and pseudo-nucleus faded quickly reaching $10^{th}$ magnitude within 10 days or so. The *Observadores_cometas* group were very active in monitoring the fade and, over the following five months, carried out more than 2600 measurements of the coma brightness using photometric apertures ranging in size from 10-60 arcsec square. The lower plot shown in Figure 1 comprises R magnitude measures of the inner 10 arcsec region of the coma clearly depicting the fading lightcurve, which appears to show anomalous brightenings at intervals of 45 ± 1 d following the initial super-outburst (Kidger, 2008).

On each occasion, the periodic enhancement in the brightness of the inner coma persisted for around 8-14 days before falling back to the underlying fade rate of the lightcurve. Each enhancement is indicative of a minor outburst of activity especially as the analysis of the relative brightness of the inner coma to the outer coma shows that on each occasion there was a step change in the degree of condensation of the inner coma as shown in the upper plot of Figure 1. This latter measure is independent of any frame-to-frame differences in magnitude calibration.

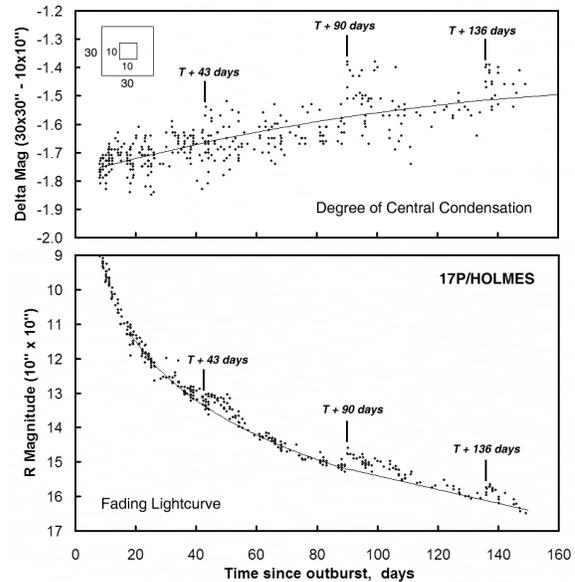

Figure 1. Photometry of comet 17P/Holmes showing the fading brightness of the inner coma (10x10 arcsec box) alongside a plot of the degree of condensation of the pseudo-nucleus (shown in the upper plot). Data courtesy of M. Campas, M.R. Kidger and the *Observadores_cometas* group.

## 4. *"Project Holmes"*

The spectacular outburst of 17P/Holmes in 2007 October drew much attention including that of the present author, who within three weeks of the outburst submitted a paper describing a possible mechanism to explain the phenomenon (Miles 2007). A key feature of the proposed mechanism is the requirement that the comet nucleus rotates unusually slowly. Indeed, the prediction made was that the rotational period of the nucleus would lie in the range, 15-75 days. This compares with known values for eleven Jupiter-family comet nuclei ranging between about 5-41 hr (Fernandez, 2005).

The photometric data discussed in the previous section appears to show a 45-day periodicity in activity, re-occurring three times after the initial super-outburst. Given that a large mass of material had been ejected by the initial explosion, it is reasonable to suppose that the site of the explosion would remain mildly active afterwards for some months at least, and that each time the incident solar radiation at the site reached a maximum, an enhancement in activity would result.

Clearly more observations would be required to further investigate 17P/Holmes but this would have to wait until late 2008 when the comet had passed solar conjunction and would reappear in the morning sky in the constellation of Cancer. Unfortunately by then, the comet would be at a heliocentric distance of about 4 AU and therefore much fainter and effectively out of range of small telescopes. The author and colleague, George Faillace, applied for and were allocated time on the 2.0-m Liverpool Telescope (LT) on La Palma, Canary Islands, to monitor 17P/Holmes during the period 2008 Oct – 2009 Jan. In addition, a proposal was submitted to the Faulkes Telescope Project to use the 2.0-m Faulkes Telescope North (FTN) instrument on Maui, Hawaii. At the time, these two telescopes were almost identical having the same optical design, camera type and Sloan filters. The objective was to monitor the comet at regular intervals during times when the bright Moon was absent from the sky taking a series of exposures during observing runs of 0.5-1.0 hr duration. In practice, the comet was imaged from the LT on 8 nights, of which 4 nights were satisfactory for photometry. From the FTN, a total of 16 nights were suitably photometric out of 27 nights in all.

## 5. Photometric Reduction – An Alternative Approach

### 5.1 Observing Methodology

The primary objective of "*Project Holmes*" is to determine the rotation period of the nucleus of 17P/Holmes. This is a tall order given that the comet has a significant motion occupying different telescopic fields on different nights. To identify a periodicity of say 45 days requires one to carry out absolute photometry of the target over an entire apparition lasting some 4-5 months. Furthermore, just to make life even more difficult, since the 2007 outburst, the comet nucleus has been surrounded by a relatively bright coma and so any rotational signature will be diluted by the coma contribution within the photometric aperture. Residual cometary activity will also complicate the resultant lightcurve. It is therefore very important to be able to accurately measure the brightness of the innermost region of the coma at frequent intervals. For these reasons, only relatively large-aperture telescopes situated at a good observing site would be suitable: hence the choice of 2.0-metre aperture instruments.

Absolute photometry requires access to suitable reference stars preferably on the same frame as the target object. It was clear that the preferred filter passband to use for all images would be Sloan-r'. Signal to noise would be maximized with this filter and comparison stars from the SDSS catalogs could be utilized for calibrating frame zero points.

Blurring of images due to motion of comets and asteroids is potentially problematic if photometry of the highest precision is sought. Various options were considered including tracking the telescope offset from sidereal rate to match the motion of the comet – this facility is available with the LT. The best solution, however, was to track the scope at sidereal rate and to select the exposure time of each frame so that the comet moved exactly 2.0 pixels (0.56 arcsec) during the time interval the shutter was open. Analysis of the cometary images was then performed using standard-sized elliptical photometric apertures to take account of this motion blur.

### 5.2 Multi-aperture Photometry of SDSS Stars in Stacked Frames

Examples of stacked frames are shown in Figure 2. Typically, sets of 5-10 frames were processed in two ways; one involved stacking on the stars (zero motion, as per the upper image) using the software, *Astrometrica*, to create a deep, high SNR image from which multi-aperture photometry was performed using circular apertures of radius, 3, 4, 5, 6, 8, 10 and 12 pixels; the other involved stacking the same set of images using *Astrometrica* but in this case the centre of each frame was moved at the same speed and in the same direction as the comet (typically in the range 0.1-0.5 arcsec/min). The latter approach yields a stacked frame as per the lower image in Figure 2 and it is this image which is used to measure the intensity of the comet as described in Section 5.3.

A total of 48 stacked sets of images were generated from 20 nights of observation on the FTN and LT between 2008 Oct 1 and 2009 Feb 26 (Table 1 in Appendix). In each case, the frame zero point was measured with the aid of the proprietary software, *AstPhot32*, written by Stefano Mottola. SDSS stars listed in Data Release 7 were selected as comparisons. They were generally chosen from the magnitude range, $14.0 < r' < 17.0$ to maximize photometric accuracy. Near the bright end of the range, care was taken to make sure that pixels within the star images were not close to saturation. For each 4.6 arcmin square field of view, some 3-10 SDSS stars were suitable for use as photometric references.

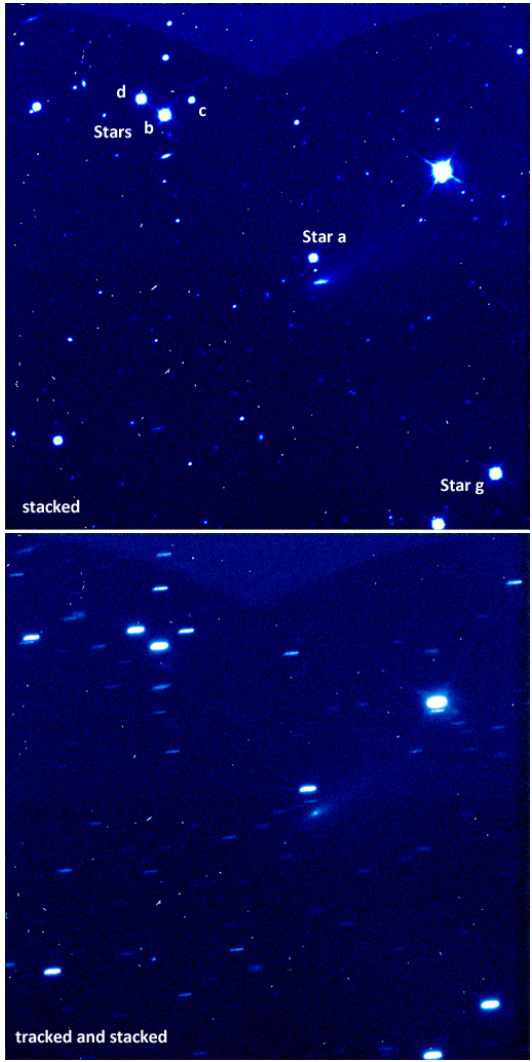

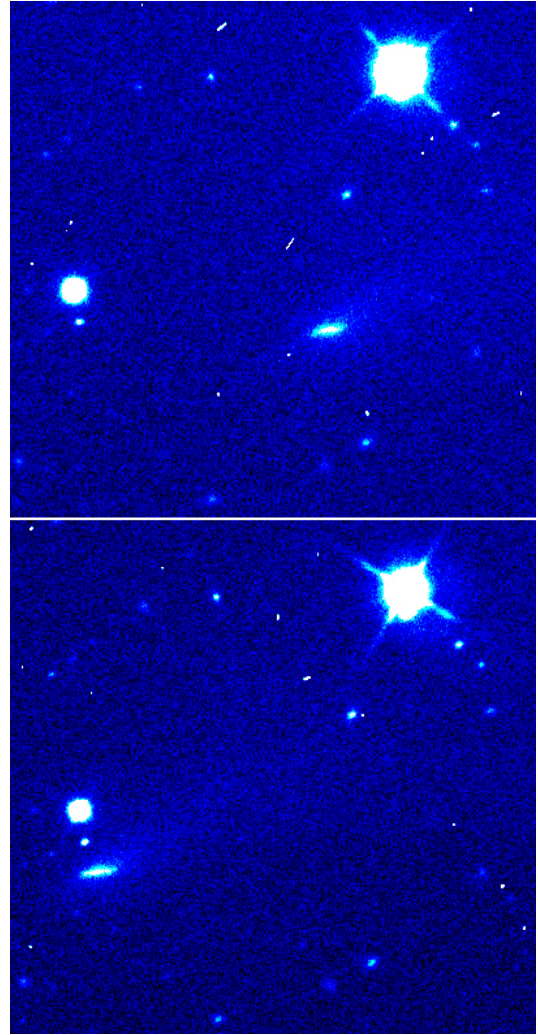

**Figure 2.** An example of stacks of 7 x 68-sec Sloan-r' filter exposures taken on 2009 Jan 20.527 (mid-time) using the 2.0-m aperture Faulkes Telescope North. In the upper image, frames were stacked on the stars to permit photometry of SDSS stars, a,b,c,d, and g enabling an accurate determination of the frame zeropoint. In the lower image, the eight frames have been stacked and tracked with the motion of the comet applied (0.492 arcsec/min at P.A. 276.0 deg).

**Figure 3a.** Magnified view of stacked frames (each 7 x 68-sec) showing a 1.8 arcmin square field traversed by the comet located near R.A. 08:54:25, Dec. +26:38:00. Mid-times: 2009 Jan 20.527 (lower) and 20.598 (upper).

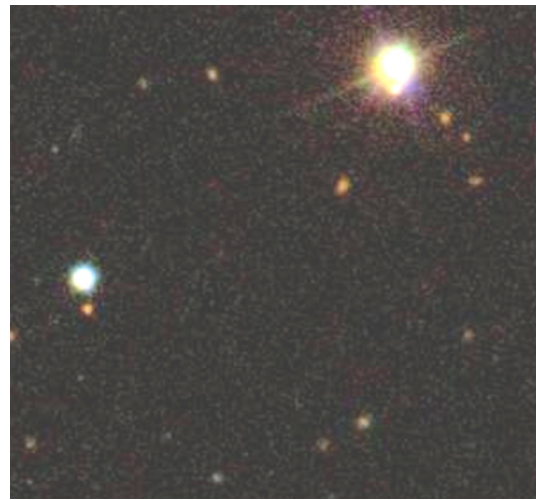

**Figure 3b.** SDSS chart of the same region of sky depicted in Figure 3a (plate centre at R.A. 08:54:25, Dec. +26:38:00).

The match found between the Sloan-r' instrumental magnitudes and standard Sloan-r' data was virtually independent of star color. The r' transformation coefficient was determined to equal -0.003 ± 0.010 as a function of g'-r'. Stars with a g'-r' color index of less than +0.3 or greater than +1.4 were nonetheless avoided. The average standard error for the frame zero point of the 48 stacks was 0.0045 mag with the calibration of the smallest (3- and 4-pixel radius) photometric apertures being systematically slightly worse than the average. This level of calibration accuracy was very encouraging and clearly al-lows the possibility of highly precise photometry even though the comet traversed some 15 degrees of sky during the course of the apparition.

During the early part of the monitoring period, the inner (3-pixel radius) coma of the comet was a little fainter than r' = 20.0 and the outer measured region (12-pixel radius) was just fainter than r' = 19.0. Clearly with the comet so faint, one of the main potential sources of photometric error would be any inadvertent inclusion of faint background field stars or more especially galaxies in the measuring aperture. Two approaches were employed to minimize the risk of contaminating the aperture: one involved checking the stacked images of the field traversed by the comet at different times as shown in Figure 3a. Here it can be seen that at both the epochs shown, the comet occupied regions of the sky devoid of other objects down to about magnitude 24. The second approach was to check the SDSS chart such as the one shown in Figure 3b ahead of time to ensure that the track of the comet was free from background objects. Indeed, in the case of the FTN, it was possible to select specific observing times to satisfy this requirement: this helped greatly in assuring that if the sky was clear for each observing run, useable images would be secured. Fortunately, the area of sky traversed by the comet lay at a relatively high galactic latitude (about +31 to +43 degrees) in the constellation of Cancer and so it was relatively easy to find uncontaminated regions of sky along the track.

### 5.3 Circular and Elliptical-aperture Photometry of the Comet Nucleus in Tracked and Stacked Frames

Photometry of the comet was undertaken on the tracked and stacked frames using *AstPhot32*. In comet photometry it is important to subtract the correct background sky signal from the coma measure. The procedure used here was to measure two diametrically opposite areas of sky located about 150 pixels (about 40 arcsec) either side of the pseudo-nucleus, which from the stacked frames could be seen to be of a uniform intensity and free of faint objects. Each area of sky contained 900 pixels and the average of these two areas then provided a suitably accurate measure of the background signal at the location of the comet.

As mentioned in Section 5.1, aperture photometry of the pseudo-nucleus was performed using elliptical apertures tilted with the long axis pointing in the direction of motion of the comet. For the large majority of frames, the exposure time corresponded to exactly 2.0 pixels motion on the CCD and so the photometric apertures employed had short- and long-axis dimensions of 8x6, 10x8, 12x10, 14x12, 16x14, 18x16, 22x20 and 26x24 pixels. These yielded the integrated intensity within an equivalent 3-, 4-, 5-, 6-, 8-, 10- and 12-pixel radius aperture. Under good sky conditions, it was possible to reach SNRs of 60-70 in tracked and stacked images of the 20$^{th}$ magnitude inner coma.

### 5.4 Normalising Photometry to Constant Seeing Conditions

Given the procedure outlined above, the main residual source of error in photometric measures of the faint inner coma of distant comets arises from variations in the FWHM or seeing. The reason for this is that differential photometry is carried out relative to stars, which are point sources. If the comet exhibited a 'bare nucleus' (i.e. no attendant coma present), then it too would be a point source and seeing would be fully corrected for. If, however, there is an appreciable coma surrounding the nucleus then, as the seeing deteriorates, an increasing proportion of the coma spreads into and contaminates the measuring aperture. The end result is that as the seeing gets worse, the apparent brightness within the measuring aperture increases.

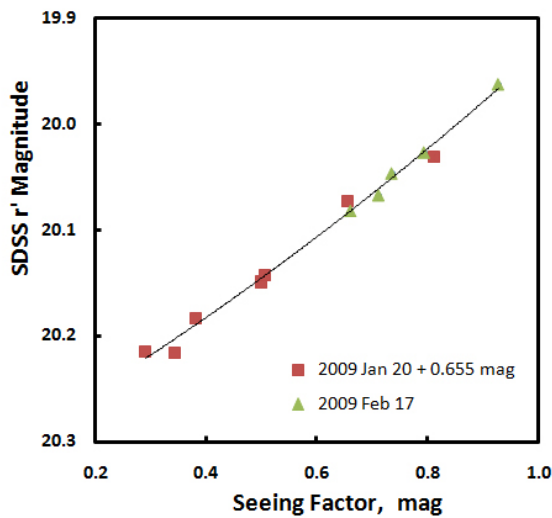

Figure 4. Effect of changes in seeing on the measured r' magnitude of 17P/Holmes (3x4-pixel elliptical aperture). The seeing factor is the average difference in the intensity of reference stars measured using a 3-pixel radius photometric aperture compared to the same stars measured using a 12-pixel aperture, expressed as magnitude.

This effect is shown in Figure 4 which illustrates the change in measured brightness of the inner coma using a 3-pixel aperture versus seeing for tracked and stacked images of the comet recorded on 2009 Jan 20 and Feb 17. The term 'seeing factor' employed here corresponds to the average difference in the intensity of reference stars measured using a 3-pixel radius photometric aperture compared to the same stars measured using a 12-pixel radius aperture, expressed as magnitude.

Given the form of the relationship plotted in Figure 4, it is possible to generate an empirically-derived correction for changes in seeing for measurements in each of the smaller photometric apertures. This was done for the four smallest apertures so that measurements could be corrected to a standard seeing factor of 0.500.

Few observers take the trouble to correct for changes in seeing when performing comet photometry and yet as can be seen here, it is a key factor if ultimately the highest precision is to be extracted from the image data. Some care has to be taken in deriving empirical corrections for seeing. For example, if the nature of the coma changes significantly then this may have an effect on the value of the correction and so the calibration may need to be checked from time to time. One is also reliant on changes in seeing taking place during the course of an observing run in order to get a handle on this effect. To maximize calibration accuracy, one trick is to stack a subset of those frames having the *poorest* seeing, and another set having the *best* seeing so as to generate datapoints over the widest possible range of seeing on a particular night. This was done in the case of the Jan 20 and Feb 17 images (shown in Figure 4) to yield the datapoints at either end of the range of seeing experienced on these two nights.

## 6. Photometric Results

### 6.1 The Fading Coma About One Year After the Initial Super-outburst

The first photometric observations were made on 2008 Oct 1 when the comet was located in the morning sky at a solar elongation of just 60 degrees. So that all data could be compared in a standard way, the aperture measurements were reduced following the approach used by other observers of distant, relatively inactive comets. A good example is that of Snodgrass et al. (2006). The principle of the data reduction method is to treat comets as though they were asteroids. So the measured magnitude, $m_r$, in any given aperture is reduced to unit heliocentric distance, $R_h$, unit geocentric distance, $\Delta$, and zero phase angle, $\alpha$, on the basis of the relationship:

$$r'(1,1,0) = m_r - 5.\log(R_h . \Delta) - \beta.\alpha \qquad (2)$$

where distances are expressed in AU, $\alpha$ in degrees, and $\beta$ is the phase coefficient in magnitudes per degree. A commonly assumed value for $\beta$ of 0.035 mag/deg was adopted for the analysis (Lamy et al., 2004).

Equation (2) is a quite precise description of the changes in brightness of low-albedo asteroids and bare cometary nuclei. With 17P/Holmes, the comet still retained a very appreciable coma and so the analysis was further extended by interpolating the reduced magnitude in different apertures to standard dimensions at the distance of the comet. Various values were examined and a volume space having a radius of 3,000 km radius was generally used for the present analysis. In practice this corresponded to a radius of 3.5-4.6 pixels depending on the Earth-comet distance at the time.

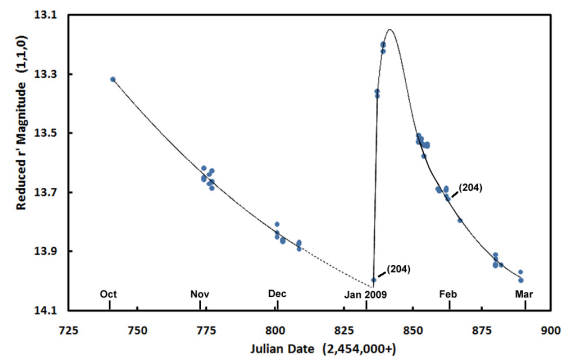

Figure 5. Lightcurve of 17P/Holmes from observations by R. Miles and G. Faillace using the 2.0-m Faulkes Telescope North, the 2.0-m Liverpool Telescope, and by L. Buzzi using the 0.6-m reflector of Schiaparelli Observatory indicated by the points marked (204). Reduced magnitude, r'(1,1,0), corresponds to a 3,000-km radius photometric aperture normalized to a seeing factor equal to 0.500 mag.

Figure 5 shows the photometric results obtained. During the last three months of 2008, the light reflected from the nucleus and inner 3,000 km radius of the coma declined by about 0.7 mag, i.e. by almost a factor of two. This decline rate is almost twice the fade rate (1.5 mag) for the inner coma (reduced to standard distance and phase angle) two to five months after the initial super-outburst (from data as shown in Figure 1, provided by M. Campas and R. Naves).

The coma was expected to continue declining in intensity so as to reveal an increasing contribution in the light reflected from the nucleus thereby making it easier to distinguish its rotational characteristics. In practice, sometime between Dec 8 and the next observations on Jan 5, the comet was found to have undergone a very marked outburst.

### 6.2 Discovery of a new outburst on 2009 Jan 4/5

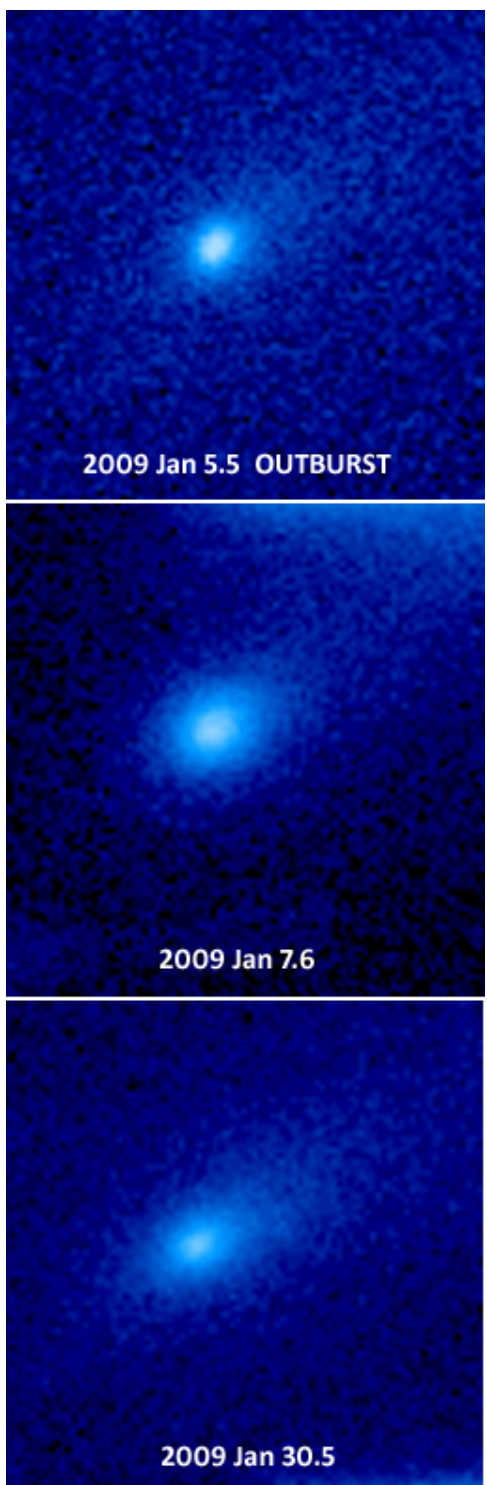

**Figure 6. Figure 6. Comet 17P/Holmes in outburst. Note the stellar appearance of the inner coma on Jan 5.5, i.e. less than 1 day after the onset of the outburst. Field of view: 24 arcsec square, log stretched images, false colour (all images: 2.0-m FTN).**

Having gone into outburst, the challenge was to determine the exact time when this began. Further observations were possible on Jan 7, which seemed to show that the inner coma had further brightened (see Figure 6) thus indicating that the start of the outburst must have been just a few days earlier.

To check whether anyone had observed 17P around these dates, the author looked at MPEC 2009-B43, the list of '*Observations of Comets*' issued by the Minor Planet Center on 2009 Jan 21. As it happened, two astrometric positions had been reported by Luca Buzzi of Schiaparelli Observatory (MPC Code 204) for Jan 4.190 and Jan 4.224. After corresponding with Luca, he sent me the FITS images taken with a 0.6-m aperture reflector stacked both on the stars and on the comet so that the photometric reduction could be carried out. From the analysis, it appeared that at the epoch of his observations no brightening greater than about 0.15-0.20 mag had occurred. Luca was able to take another series of images of 17P on Jan 31.1 such that the change in brightness between Jan 4 and Jan 31 could be determined with a precision of about ±0.06 mag. Having observations of 17P made with the FTN on Jan 30.5, it was possible to plot the MPC 204 magnitudes on the overall lightcurve shown in Figure 6 so that the Jan 31 data fitted the lightcurve exactly thereby deriving a more accurate reduced magnitude for the Jan 4.2 measurement. The conclusion arrived at was that the start of this new outburst took place on 2009 Jan 4.7±0.5, which corresponds to 439.0 ± 0.7 days since the initial super-outburst.

### 6.3 Effect of the Outburst on the Intensity Distribution Within the Coma

In addition to the overt brightening of the comet, the effect of the new outburst was to temporarily change the degree of condensation of the pseudo-nucleus, more specifically the relative intensity of the coma in photometric apertures of increasing size. Figure 7 illustrates this as plots of the brightness in each aperture relative to the brightness in the largest photometric aperture (12-pixel radius) on different nights. The intensity distribution on 2009 Jan 30 (26 days after the outburst) was virtually identical to its pre-outburst form as recorded on 2008 Dec 8. In contrast, those on Jan 5.5 and Jan 7.6, about 1 and 3 days after the new outburst, both show a significant enhancement in the brightness of the inner coma. After about 1 day, the inner (3-pixel) coma had brightened by some 0.7 mag whereas in the largest (12-pixel) aperture, the total brightening was only 0.3 mag. At this time, 3 pixels equated to almost exactly 2,000 km in the plane of the sky at the distance of the comet (3.28 AU from the Earth). It appears that a significant proportion of material expelled during the outburst travelled away from the nucleus at speeds of less than 50 m/s.

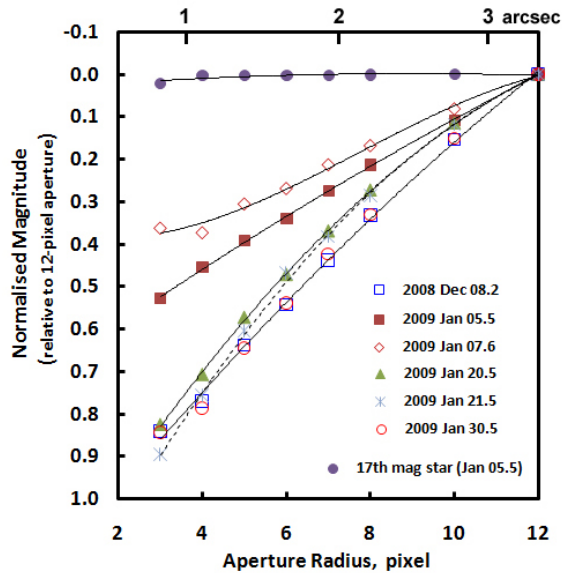

**Figure 7.** Intensity of the coma of 17P/Holmes in apertures of increasing size relative to its brightness in the largest photometric aperture (12-pixel radius). Note the enhancement in the intensity of the innermost coma on Jan 5.5 and particularly on Jan 7.6 following the outburst on Jan 4.

### 6.4 Comparisons between the 2009 Jan 4/5 outburst and minor outbursts reported after the initial super-outburst

The magnitude of the Jan 4/5 outburst attained a maximum of 0.85 ± 0.1 mag at 3,000 km radius and 0.6 ± 0.1 mag at 8,000 km radius. The decline from maximum became significant about 15 days following the start of the outburst and continued monotonically for the remainder of the observing period. This behavior is characteristic in degree and duration to the suspected three minor outbursts described in Section 3.3, which were indicative of a 45±1 day periodicity. The tantalizing question is whether outburst activity continues to arise intermittently from a particular region of the nucleus such that the interval of 439 days reported here for the second significant outburst of 17P/Holmes corresponds to exactly TEN revolutions of the nucleus and hence its rotation rate is actually 44 days.

### 6.5 Photometric Scatter and Constraints on the Rotational Characteristics of the Nucleus

Comet nuclei exhibit a wide range of rotation periods, the large majority of which are less than 36 hours. They also tend to be relatively elongated in comparison with asteroids in that their projected axial ratios are typically in the range, 1.3-2.0 (Snodgrass et al., 2006). Since elongated bodies generally exhibit two maxima and two minima in their lightcurve per 360° rotation, it is instructive to examine the photometry reported here to test whether periodicity is apparent over short time-scales.

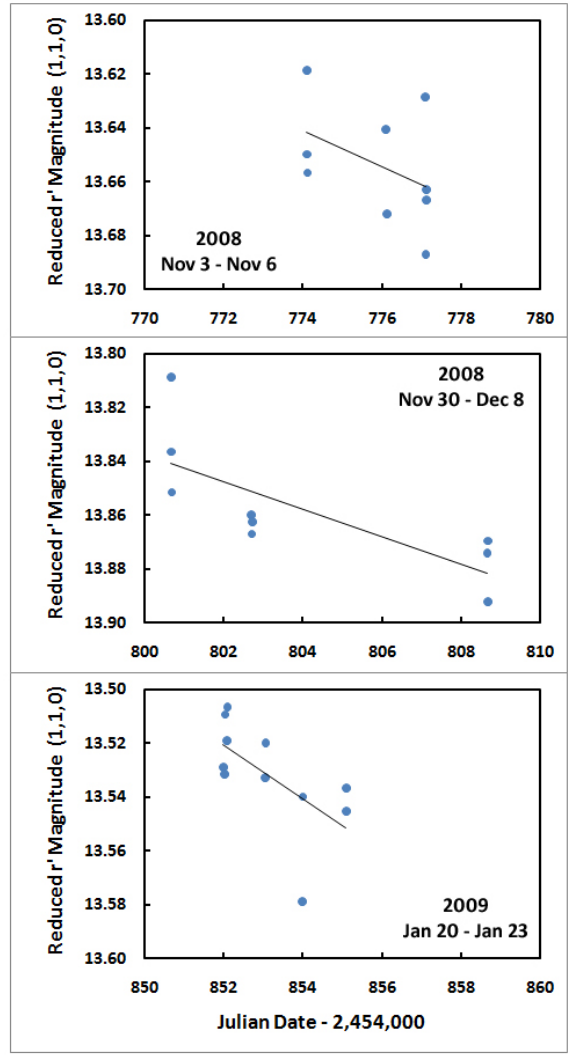

**Figure 8.** Detail of the fading lightcurve of 17P/Holmes at three epochs. Middle plot derived from images taken with the LT, the others using FTN images. Lines shown were obtained by 'least squares' linear regression.

Figure 8 depicts three sections of the lightcurve each plotted to the same scale. The standard deviation of the points relative to a least-squares linear regression is just 0.015-0.019 mag, so that if the rotation rate were less than 36 hours, the peak-to-trough amplitude of the combined light from the nucleus *and* coma would need to be less than about 0.03 mag.

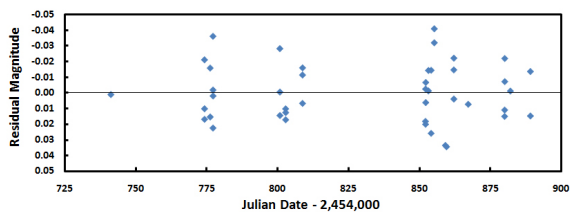

**Figure 9.** Plot of residual magnitude vs. Julian Date. The residuals are in the sense 'observed' minus 'calculated' (the latter using two separate 2nd order polynomial fits to the data between JD2454741-809 and 2454852-889).

A further analysis was carried out using the period-search software, *Peranso* written by Tonny Vanmunster. Here, two separate 2nd-order polynomial fits were made to the monotonic fades in the lightcurve before and after the outburst: that is between JD2454741-809 and between JD2454852-889. The difference between the measured and calculated magnitudes is plotted in Figure 9 against Julian Date. The ANOVA method was then used within *Peranso* to search the residual magnitude data for periods of less than 36 hours but nothing of significance was found.

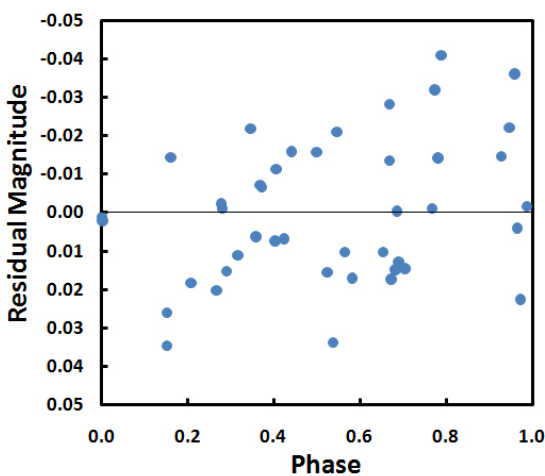

**Figure 10.** Example of a phase plot of residual magnitudes folded on a trial period of 16.3 hr. No significant periodicity is evident.

An example of the 'best' phase plot that could be generated is that shown in Figure 10, where the data have been folded on a trial period of 16.3 hr. No significant periodicity is evident. Furthermore, the amplitude of any variability if present is hidden by the scatter in the data, hence any short time-scale periodic variability is probably less than 0.02 mag in amplitude.

Snodgrass et al. (2006) were able to observe 17P/Holmes in 2005 when it was returning from aphelion at a heliocentric distance of 4.66 AU at which time no coma was present as seen using a 3.5-m aperture telescope. They determined an average R magnitude of 22.86 ± 0.02 at 2005 Mar 6 and a V-I color index of 0.85. This observation equates to an R(1,1,0) magnitude of 16.22 or an r'(1,1,0) magnitude of 16.4 given the somewhat different passbands of the filters used. Since the r'(1,1,0) magnitude during the fading sections of the lightcurve (Figure 5) was 13.5-14.0, we can expect the contribution of the light reflected from the nucleus to amount to some 7-11% of the total light reflected within the central 3,000 km radius region.

If the short time-scale variability is less than 0.02 mag amplitude, then either the nucleus of 17P must be relatively spherical (projected axial ratio of <1.25), or, if its shape is more typical of other comet nuclei, it must have a rotational period in excess of 10 days. The remaining unlikely possibility is that the nucleus was orientated almost pole-on to the Earth.

## 7. Observations at Future Apparitions

The next few apparitions of 17P/Holmes should be favourable ones for carrying out further studies of its nucleus. The next opposition occurs on 2010 Feb 24 at R.A. 10h 25m, Dec. +05 44', i.e. in a not-too-crowded region of the sky at a galactic latitude of +49 degrees in the constellation of Sextans. According to the JPL Horizons ephemeris, at the 2010 opposition it is expected to be 1.3 mag fainter than this year's opposition and so should still be within range of a 2.0-m telescope. Opposition in 2011 is on Mar 20 at about R.A. 11h 38m, Dec. -11 58' at a galactic latitude of +47 deg in the constellation of Crater with the comet similar in brightness to the 2010 opposition. Faint background galaxies will be the most likely source of photometric error during both these apparitions.

## 8. Discussion and Conclusions

During this observing campaign, the coma associated with the nucleus of Comet 17P/Holmes has prevented a clear view of the bare nucleus. Not only did the coma fail to dissipate sufficiently quickly, it actually underwent a re-outburst further obscuring the nucleus. As such, only limited constraints have been placed on its shape and rotational characteristics which are indicative of the nucleus being a very slow rotator. Although rather speculative, the time of re-outburst has been accurately determined and this fact along with the observation by others of recurrent activity following the initial super-outburst, leads to the intriguing possibility that the nucleus spins once every 44 days.

Comets look to have different spin characteristics to asteroids in that they seem to have a flat spin rate distribution compared to the Maxwellian-shaped distribution of asteroid spin rates (Snodgrass, 2006). In simple terms, this is thought to arise from the outgassing and outburst activity of comet nuclei which serve to speed up or slow down their rotation. Hence, we can expect a significantly larger fraction of comet nuclei to be slow rotators in comparison with asteroids in line with the findings reported here.

Despite the limited conclusions arrived, this work has demonstrated a practical methodology utilizing the SDSS Catalog for accurately quantifying the brightness of short-period comets far from perihelion especially when they are still associated with a residual coma.

Whether the injection of dust into the coma is a continual one for more distant comets approaching perihelion is not certain. The application of the 'Afrho' parameter as a measure of dust production rate could be misleading in these cases. The results reported here show that, apart from the single outburst event, the nucleus faded in a uniform manner. This fact might be explained on the basis of a wide distribution of velocities within the dust coma and no replenishment of dust particles through continued outgassing from the nucleus. The reduced solar wind at larger heliocentric distance would then result in a slower dissipation of the coma.

## 9. Acknowledgements

I would like to express my sincerest thanks to my colleague, George Faillace, who has provided much inspiration, stimulated discussion and helped me persevere with this investigation. George has operated the Faulkes Telescope North on a good number of occasions and assisted in the application for time on the Liverpool Telescope.

The people at the Faulkes Telescope team based at Cardiff University, including project director Paul Roche, have been enormously helpful. The Faulkes Telescope Project is part of the Las Cumbres Observatory Global Telescope Network (LCOGTN) which provides access to a global network of robotic telescopes and supplies free resources for science education.

For several years now I have used the photometry software, *AstPhot32* written by Stefano Mottola of the DLR – Institute of Planetary Research, Berlin, for which I owe him a debt of gratitude. This software, though used manually, enables precise data to be extracted from FITS images of moving objects.

I would also like to acknowledge the help and encouragement of Mark Kidger of the European Space Astronomy Centre (ESAC), Madrid, as well as help from Montse Campas and Ramon Naves, who are key players within the *Observadores_cometas* group.

Time on the Liverpool Telescope was granted under proposal, PL/08B/14 and assistance in planning observations was provided by Chris Moss as Liverpool Telescope support astronomer. The Liverpool Telescope is operated on the island of La Palma by Liverpool John Moores University in the Spanish Observatorio del Roque de los Muchachos of the Instituto de Astrofisica de Canarias with financial support from the UK Science and Technology Facilities Council.

## 11. Appendix: Observational Details

| Date (UT) | Integration Time (sec) | Seeing factor mag | SDSS ref. stars | Zeropoint error mag | Hel. Dist. AU | Geo. Dist. AU | Phase Angle deg | r' (1,1,0) Reduced mag |
|---|---|---|---|---|---|---|---|---|
| Oct 1.613 | 960 | 0.452 | 3 | 0.011 | 3.848 | 4.244 | 13.1 | 13.319 |
| Nov 3.592 | 725 | 0.297 | 4 | 0.003 | 3.965 | 3.880 | 14.5 | 13.618 |
| Nov 3.604 | 725 | 0.248 | 3 | 0.002 | 3.965 | 3.880 | 14.5 | 13.650 |
| Nov 3.616 | 725 | 0.277 | 3 | 0.003 | 3.965 | 3.880 | 14.5 | 13.657 |
| Nov 5.598 | 1092 | 0.433 | 5 | 0.004 | 3.972 | 3.856 | 14.5 | 13.640 |
| Nov 5.615 | 936 | 0.399 | 5 | 0.004 | 3.972 | 3.856 | 14.5 | 13.672 |
| Nov 6.590 | 656 | 0.351 | 6 | 0.003 | 3.976 | 3.844 | 14.4 | 13.628 |
| Nov 6.599 | 656 | 0.262 | 6 | 0.003 | 3.976 | 3.844 | 14.4 | 13.687 |
| Nov 6.609 | 656 | 0.300 | 6 | 0.003 | 3.976 | 3.844 | 14.4 | 13.663 |
| Nov 6.619 | 656 | 0.237 | 6 | 0.003 | 3.976 | 3.844 | 14.4 | 13.667 |
| *Nov 30.170* | *1000* | *0.671* | *4* | *0.005* | *4.056* | *3.568* | *12.9* | *13.809* |
| *Nov 30.182* | *1000* | *0.705* | *4* | *0.005* | *4.056* | *3.568* | *12.9* | *13.836* |
| *Nov 30.194* | *1000* | *0.676* | *4* | *0.005* | *4.056* | *3.568* | *12.9* | *13.851* |
| *Dec 2.198* | *1000* | *0.473* | *4* | *0.005* | *4.063* | *3.546* | *12.7* | *13.860* |
| *Dec 2.210* | *1000* | *0.425* | *4* | *0.004* | *4.063* | *3.546* | *12.7* | *13.867* |
| *Dec 2.222* | *1000* | *0.440* | *4* | *0.004* | *4.063* | *3.546* | *12.7* | *13.862* |
| *Dec 8.143* | *1000* | *0.684* | *8* | *0.007* | *4.083* | *3.484* | *11.9* | *13.874* |
| *Dec 8.155* | *1000* | *0.631* | *8* | *0.006* | *4.083* | *3.484* | *11.9* | *13.892* |
| *Dec 8.167* | *1000* | *0.642* | *8* | *0.006* | *4.083* | *3.484* | *11.9* | *13.869* |
| Jan 5.514 | 954 | 0.679 | 6 | 0.004 | 4.175 | 3.279 | 6.3 | 13.360 |
| Jan 5.530 | 954 | 0.696 | 5 | 0.002 | 4.175 | 3.279 | 6.3 | 13.375 |
| Jan 7.570 | 770 | 1.218 | 5 | 0.003 | 4.181 | 3.271 | 5.8 | 13.199 |
| Jan 7.584 | 770 | 1.212 | 3 | 0.004 | 4.181 | 3.271 | 5.8 | 13.224 |
| Jan 7.597 | 616 | 1.048 | 3 | 0.005 | 4.181 | 3.271 | 5.8 | 13.205 |
| Jan 20.487 | 544 | 0.499 | 5 | 0.005 | 4.222 | 3.254 | 2.8 | 13.529 |
| Jan 20.527 | 476 | 0.343 | 5 | 0.004 | 4.222 | 3.254 | 2.8 | 13.532 |
| Jan 20.535 | 476 | 0.382 | 5 | 0.003 | 4.222 | 3.254 | 2.8 | 13.509 |
| Jan 20.589 | 476 | 0.655 | 6 | 0.002 | 4.222 | 3.245 | 2.8 | 13.519 |
| Jan 20.598 | 476 | 0.506 | 8 | 0.002 | 4.222 | 3.245 | 2.8 | 13.507 |
| Jan 21.547 | 476 | 0.357 | 8 | 0.003 | 4.225 | 3.255 | 2.6 | 13.533 |
| Jan 21.562 | 476 | 0.356 | 6 | 0.008 | 4.225 | 3.255 | 2.6 | 13.520 |
| Jan 22.487 | 405 | 0.475 | 6 | 0.008 | 4.228 | 3.256 | 2.4 | 13.579 |
| Jan 22.493 | 472.5 | 0.596 | 9 | 0.004 | 4.228 | 3.256 | 2.4 | 13.540 |
| Jan 23.589 | 476 | 0.770 | 9 | 0.004 | 4.231 | 3.258 | 2.4 | 13.545 |
| Jan 23.599 | 476 | 0.877 | 8 | 0.002 | 4.231 | 3.258 | 2.4 | 13.537 |
| Jan 27.505 | 336.5 | 0.551 | 8 | 0.002 | 4.243 | 3.267 | 2.0 | 13.688 |
| *Jan 27.922* | *720* | *0.500* | *6* | *0.004* | *4.245* | *3.268* | *2.0* | *13.697* |
| Jan 30.487 | 680 | 0.619 | 6 | 0.007 | 4.253 | 3.277 | 2.2 | 13.693 |
| Jan 30.500 | 680 | 0.699 | 8 | 0.002 | 4.253 | 3.277 | 2.2 | 13.686 |
| Jan 30.513 | 680 | 0.642 | 8 | 0.002 | 4.253 | 3.277 | 2.2 | 13.712 |
| Feb 4.565 | 140 | 0.707 | 5 | 0.007 | 4.268 | 3.301 | 3.0 | 13.795 |
| Feb 17.396 | 738 | 0.711 | 4 | 0.005 | 4.306 | 3.398 | 5.8 | 13.947 |
| Feb 17.414 | 820 | 0.793 | 5 | 0.008 | 4.306 | 3.398 | 5.8 | 13.943 |
| Feb 17.434 | 820 | 0.927 | 7 | 0.007 | 4.306 | 3.398 | 5.8 | 13.911 |
| Feb 17.449 | 880 | 0.735 | 7 | 0.008 | 4.306 | 3.398 | 5.8 | 13.926 |
| Feb 19.428 | 924 | 0.525 | 5 | 0.004 | 4.312 | 3.417 | 6.3 | 13.946 |
| Feb 26.485 | 490 | 0.553 | 10 | 0.003 | 4.333 | 3.494 | 7.8 | 13.970 |
| Feb 26.494 | 550 | 0.517 | 10 | 0.003 | 4.333 | 3.494 | 7.8 | 13.998 |

**Table I. Observing log for 17P/Holmes carried out using the Faulkes Telescope North and the Liverpool Telescope (entries in italics). Reduced magnitude, r'(1,1,0), corresponds to a 3,000-km radius photometric aperture normalized to an intermediate seeing factor of 0.500 mag.**